\newif\iffigs\figsfalse
\figstrue
\catcode`\@=11
\def\slash{\mathpalette\make@slash}
\def\make@slash#1#2{\setbox\z@\hbox{$#1#2$}%
\hbox to 0pt{\hss$#1/$\hss\kern-\wd0}\box0}
\catcode`\@=12 
\def\d{\hbox{d}}

\def\be{\begin{equation}}
\def\ee{\end{equation}}
\def\bea{\begin{eqnarray}}
\def\eea{\end{eqnarray}}
\def\es{\!=\!}
\def\I{\hbox{Im}}

\def\pa{\partial}
\def\es{\!=\!}
\def\ha{{1\over 2}}
\def\al{\alpha}
\def\>{\rangle}
\def\<{\langle}
\def\mtx#1{\quad\hbox{{#1}}\quad}

\def\di{D\!\!\!\!/}
\def\nn{\nonumber}

\def\tr{\hbox{tr}}
\newcommand{\refs}[1]{(\ref{#1})}

\documentstyle[aps,epsf]{revtex}
\begin{document}

\title{  SU(2) Yang-Mills Theory with extended Supersymmetry in a 
Background Magnetic Field}

\author{D.G.C. McKeon}
\address{Department of Applied Mathematics,\\ 
         University of Western Ontario, 
         London, ON N6A 5B7,
         Canada}

\author{I. Sachs}
\address{Department of Mathematical Sciences,\\
	 University of Durham, Durham DH1 3LE, UK}

\author{I.A.~Shovkovy\thanks{On leave of absence from 
Bogolyubov Institute for Theoretical Physics, 252143, 
Kiev, Ukraine}}
\address{Physics Department,\\ University of Cincinnati,
Cincinnati, OH 45221-0011, USA}

\maketitle

\vspace{1.5cm}

\widetext
\begin{abstract}     
The vacuum structure of $N\es 2$ 
(and $N\es 4$) supersymmetric Yang-Mills theory is analyzed in detail 
by considering the effective potential for constant background scalar- 
magnetic fields within different approximations.  
We compare the one-loop approximation with- or without instanton improved 
effective coupling with the one-loop result in the dual desription. 
For $N\es 2$ we find that non-perturbative monopole 
degrees of freedom remove the non-trivial minima present in the 
(improved) one-loop potential in the strong-coupling regime. The combination 
of Yang-Mills and dual desription leads to a self-consistent effective 
potential over the full range of background fields. 

\end{abstract}




\section{Introduction}
\label{one}

Much attention has recently been focused on $N\es 2$ supersymmetric vector
theories \cite{SW,SW2}. A set of inequivalent vacuum states exists in
these models, as the classical potential is proportional to $Tr([\phi^*,
\phi]^2)$; distinct supersymmetric--invariant, zero--energy vacuum states
are parameterized by the constant scalar component $\phi(x)$ taking its
value in the Cartan subalgebra of the gauge group.  In \cite{SW} it has
been argued that the low energy physics of the strong-coupling regime of
these theories is equivalently described by a weakly coupled dual theory.
The analysis in \cite{SW} was restricted to the vacuum manifold. Attempts
to generalize the duality away from the vacuum have been presented in
\cite{Matone,FS,Yung}, taking into account higher derivative terms in the
effective action. In this paper, we propose another step in this direction
by considering a constant Abelian background field strength and a constant 
scalar field, aligned in the same direction in group space. We compute the 
one--loop effective potential for $N\es2$
and $N\es4$ theories, using techniques similar to those used in \cite{SS}
where the effect of an external magnetic field on the symmetry breaking
patterns in a non--Abelian Higgs model was examined. For $N=4$ the
one--loop effective potential should be reliable in the context of
perturbation theory, as the coupling constant, once chosen to be small, is
not affected by radiative corrections. In the case of $N\es 2$ we improve
the one--loop calculation with the exact results \cite{SW}, therefore
including all perturbative and non-perturbative contributions up 
to second order in the external magnetic field. 

For a given non-zero magnetic field the classical vacuum degeneracy for
the scalar field is lifted by quantum corrections. More precisely, the
effective potential has a relative minimum at $|\phi|^2/|B|=O(1)$, where
$\phi$ and $B$ denote the background scalar and magnetic fields
respectively. While the lifting of the classical degeneracy for the scalar
field is expected, we can only determine the value of the scalar field in
terms of the background magnetic field, and for $N\es 2$ we find that both
the one--loop and the instanton improved one--loop effective potential
have a minimum at a non-vanishing value of the external magnetic field for
$B$.  At the one--loop level $B_{min}$ is of the order of
$\Lambda^2_{QCD}$, where the running coupling is large and the one--loop
approximation therefore is not reliable. Ideally we should therefore
include higher order as well as non-perturbative contributions for the 
higher orders in the magnetic field as well. This, of
course, is beyond reach at present. However, the dual theory which should
describe the the strong-coupling low energy physics of this model
\cite{SW} is weakly coupled in this regime and a one--loop calculation
within the dual model should be reliable. Note that the effective 
potential, unlike the effective action, has a physical interpretation and 
should therefore be duality invariant. If duality is realized for at
least a small but non-vanishing magnetic field then the one-loop
calculation in the dual theory should contain all relevant non-perturbative
corrections in the original formulation. This is reasonable, although the
duality conjecture has been proved only in the zero-energy limit
\cite{USW} (see however \cite{Matone,FS,Yung}). We take this as motivation
for the assumption that the strong-coupling effective potential be
approximated by the (improved) one--loop effective potential of the dual
theory which is $N\es2$ supersymmetric $QED$ \cite{SW}. We find that the 
non-trivial minima are indeed removed by the monopole dynamics 
as described by the the dual action. As a result the combination of 
Yang-Mills- and dual description leads to to a self consistent effective 
potential over the full range of background fields. 

In either formulation the effective potential has a non-vanishing 
imaginary part.
While this imaginary part is normally associated with unstable (tachyonic)
modes \cite{BMS,Tseytlin1}, we can argue that they are eliminated by non-perturbative effects
as in $QCD$ \cite{NN,F,L,C}. 
In the present case, it may be interpreted as arising from monopole
production in the presence of an external magnetic field. 

The paper is organized as follows: 
In the next section we compute the one--loop effective potential for
$N\es2$ super Yang-Mills (SYM$_2$) theory for the 
background field configuration
described above. Section three deals with the actual computation of the
functional determinants arising in the one--loop computation of the
effective potential and examines the vacuum structure. We repeat this for
$N\es4$ super Yang-Mills (SYM$_4$) theory in section four. In section five we
improve the $N\es2$ potential by including all instanton corrections to
the scalar field dependence of the effective coupling. The corresponding
effective potential is evaluated numerically. Next we obtain the effective
potential to one--loop order in the dual theory ($N\es 2$ super $QED$)
using the background fields dual to those above.  The structure of this
dual effective potential is determined numerically and compared with that
of the original model. 
Section six contains our conclusions. The computation of
the functional determinant for a general electromagnetic field is
explained in the appendix.


\section{The N=2 Model}
\label{two}
A harmonic superspace formulation of SYM$_2$ was presented in 
\cite{9703147} where furthermore the non-renormalisation 
theorems for SYM$_2$ were revisited within that framework. For the finite 
contributions to the effective action we find it however easier 
to work in component formulation. The action is then given by
\begin{eqnarray}
S&=&\int d^4 x \left\{-\frac{1}{4g^2}F^a_{\mu\nu}F^{a\mu\nu}
-\frac{\theta}{32\pi^2}F_{\mu\nu}^a\tilde
F^{a\mu\nu}-(D_\mu\phi)^{*a}(D^\mu\phi)^{a}
-\bar{\chi}\not{\!\!D}\chi\right. 
\nonumber \\   
&+&\left.
\frac{1}{2}g^2(f^{abc}\phi^b\phi^{*c})^2
+\frac{ig}{\sqrt{2}}f^{abc}
  [\bar{\chi}^{a}\gamma_{-}\chi^{c}\phi^{b}
    +\bar{\chi}^{a}\gamma_{+}\chi^{c}\phi^{*b}] \right\}, \label{1}
\end{eqnarray} 
where $\gamma_{\pm}=1\pm\gamma_{5}$,
$\{\gamma_{\mu},\gamma_{\nu}\}=-2g_{\mu\nu}$ and $g_{\mu\nu}=\hbox{diag}(-+++)$
as in \cite{VMNP}. We take the gauge group to be $SU(2)$ and we align the
background fields so that 
\begin{equation}
\phi^{a}(x)= f \delta^{a3}+ h^{a}(x) \mtx{and} 
A_{\mu}^{a}(x)=-\frac{1}{2}F_{\mu\nu} x^{\nu} \delta^{a3} + 
Q_{\mu}^{a}(x),\nonumber
\end{equation}
with $f$ and $F_{\mu\nu}$ constant. 
The gauge fixing Lagrangian is taken to be a modified 
version of the $R_{\xi}$ gauge \cite{FLS}, 
\begin{eqnarray}
{\cal L}_{gf}=-\frac{1}{\xi g^2}\left[
   \frac{1}{2} (\partial_\mu Q^{\mu})^2
+ \Bigg(
(\partial-iA)^{\nu}Q_{\nu}^{+}
+ig^2\xi(f^{*}h^{+}+fh^{*+}) 
\Bigg)
\Bigg( 
(\partial+iA)^{\mu}Q_{\mu}^{-}
-ig^2\xi(f^{*}h^{-}+fh^{*-})\Bigg)\right],
\label{3}
\end{eqnarray}
where
\begin{eqnarray}
A_{\mu}=-\frac{1}{2}F_{\mu\nu} x^{\nu},\quad 
h^{\pm}=\frac{h^1\mp ih^2}{\sqrt{2}} ,\quad  
h^{*\pm}=\frac{h^{*1}\mp ih^{*2}}{\sqrt{2}},\quad 
Q=Q^3,\quad 
Q^{\pm}=\frac{Q^1\mp iQ^2}{\sqrt{2}}.
\end{eqnarray}
The Faddeev--Popov ghost Lagrangian associated with this 
gauge fixing is, to leading order in the quantum fields when $\xi=1$,
\begin{eqnarray}
{\cal L}_{FP}=\bar{c_1}[(\partial+iA)^2-2g^2f^*f] c_1
             +\bar{c_2}[(\partial-iA)^2-2g^2f^*f] c_2,
\label{4}
\end{eqnarray}
while the one--loop contributions arising from ${\cal L}+{\cal L}_{gf}$ 
are of the form 
\begin{eqnarray}
&&Q^+_\mu (\Delta^{-1})^{\mu\nu}Q^-_\nu+ 
\left(\begin{array}{ll}
h^{*+} & h^{+}\end{array}                                                  %
  \right) \left(\begin{array}{ll}0 
              & \Delta^{-1}_{0}\\
\Delta^{-1}_{0}
               & 0\end{array}
  \right)\left(\begin{array}{l}h^{*-} \\ h^{-}\end{array}\right)
+\left(\begin{array}{ll}\bar{\chi}^{+} & (\chi^{+})^T 
    \end{array}                                                  %
  \right)
\left(\begin{array}{ll}0 
                  & \Delta_{1/2}^{-1}\\
(\Delta_{1/2}^{-1})^T
                    & 0 
       \end{array}
\right)
\left( \begin{array}{l}(\bar{\chi}^{-})^T \\ \chi^{-} 
       \end{array}
\right),
\label{5} 
\end{eqnarray}
where
\begin{equation}
\Delta^{-1}_{0}=D_{+}^{2}-M^{2},\qquad(\Delta^{-1})^{\mu\nu}=g^{\mu\nu}(D_{+}^{2}-M^{2})+2iF^{\mu\nu}\mtx{and}\Delta_{1/2}^{-1}=-i\not{\!\!D}_{+} -\frac{g}{\sqrt{2}}
(\gamma_{+}f^{*}+\gamma_{-}f),
\label{7a}
\end{equation}
respectively. In \refs{5}, ``T" refers 
to a transpose in the Dirac indices only, 
$D_{\pm\mu}\equiv \partial_{\mu} \pm i A_{\mu}$
and $M^2=2g^2f^*f$. From (\ref{4}-\ref{7a}) it is then easy to see that 
the ghost and scalar loops cancel, so that 
\begin{eqnarray}
i W^{(1)}&=&-tr \ln
\left[\Delta^{\mu\nu}\right]
+2tr\ln\left[ \Delta_{1/2} \right],
\label{7}
\end{eqnarray}
where, in addition to (\ref{7a}), we have also used
the notation
\begin{equation}
\Delta_{1/4}=\left(D_{+}^{2} + \frac{1}{2}
\sigma_{\mu\nu}F^{\mu\nu}-M^{2}\right)^{-1},
\quad \sigma_{\mu\nu} 
\equiv \frac{i}{2}[\gamma_{\mu},\gamma_{\nu}].
\label{10c}
\end{equation}

If we now regulate the logarithm and reciprocal of operators 
occurring in (\ref{7}) using $\zeta$--regularization \cite{SS,H}, 
and its generalization, operator regularization \cite{MS}, 
then we find that for operators $H_i$,
\begin{eqnarray}
\ln(H_i/\mu^2)=-\left.\frac{d}{ds}\right|_{0}\frac{\mu^{2s}}{\Gamma(s)}
\int\limits_{0}^{\infty} d (it) (it)^{s-1} e^{-iH_it},
\label{8}
\end{eqnarray}
and
\begin{eqnarray}
&&(H_1 H_2 \ldots H_N)^{-1}=\left.\frac{d}{ds}\right|_{0}
\left\{
\frac{\mu^{2s}}{\Gamma(s+1)}
\int\limits_{0}^{\infty} d (it_1) (it_1)^{s} e^{-iH_1t_1}
\ldots
\frac{\mu^{2s}}{\Gamma(s+1)}
\int\limits_{0}^{\infty} d (it_N) (it_N)^{s} e^{-iH_Nt_N}
\right\}.\nonumber\\
&&\label{9}
\end{eqnarray}
In (\ref{8}) and (\ref{9}) $\mu^2$ is an arbitrary dimensionful 
parameter.


\section{The Case of the Background Magnetic Field}
\label{three}

We now specialize to the case where $F_{\mu\nu}$ corresponds 
to a magnetic field only, so that $F_{12}=-F_{21}=B$ 
(the formulae for the most general case of the 
electromagnetic field are given in Appendix A). As has 
been shown in \cite{VMNP,Sch,S} (see also equation (\ref{A1}) 
with $K_{-}=B$ and $K_{+}=0$ in Appendix A), in this case we 
have
\begin{eqnarray}
\langle x^{\prime} | \exp ( iD_{+}^{2}t ) | x \rangle
=\frac{-i}{(4\pi t)^2} \frac{Bt}{\sin(Bt)}
\exp \left[ \frac{ix^2_{\parallel}}{4t} +
           \frac{ix^2_{\perp}B} {4\tan(Bt)} 
    + \frac{iB}{2} (x_2x_1^{\prime}-x_1x_2^{\prime}) \right],
\label{10}
\end{eqnarray}
($x^2_{\perp}\equiv(x_1-x_1^{\prime})^2+(x_2-x_2^{\prime})^2$,  
$x^2_{\parallel}\equiv(x_3-x_3^{\prime})^2-(x_0-x_0^{\prime})^2$).

Furthermore, since $(\sigma_{12})^2=1\!\!1_4$, it is easily shown that 
\begin{mathletters}
\begin{equation}
tr \exp \left[ -\frac{i}{2}\sigma_{\mu\nu}F^{\mu\nu}t\right]
=4\cos(Bt),
\label{11}
\end{equation}
and 
\begin{equation}
tr \exp \left[-2F^{\mu\nu}t\right]= 4\cos^2(Bt).
\label{12}
\end{equation}
\end{mathletters}

Since
\begin{eqnarray}
tr\ln \left[-i\not{\!\!D}_{+} -\frac{g}{\sqrt{2}}
(\gamma_{+}f^{*}+\gamma_{-}f)
\right] &=&tr\ln \left[-i\not{\!\!D}_{+} 
+\frac{g}{\sqrt{2}}(\gamma_{+}f+\gamma_{-}f^{*})
\right]
\nonumber\\
&=&\frac{1}{2} tr\ln 
\left[D_{+}^2 - M^2 + \frac{1}{2}
\sigma_{\mu\nu}F^{\mu\nu}
\right],
\label{13}
\end{eqnarray}
we find  that for the leading terms in (\ref{7})
\begin{eqnarray}
iW_{eff}&\equiv& \label{-16}
-tr \ln \left[ \Delta^{\mu\nu} \right]
+2tr \ln \left[  \Delta_{1/2}   \right]
= - tr \ln \left[ \Delta^{\mu\nu} \right]
+ tr \ln \left[  \Delta_{1/4}   \right]
\equiv i \int d^4 x {\cal L}_{eff},
\end{eqnarray}
and (\ref{8}) and (\ref{10})--(\ref{12}) imply
\begin{eqnarray} 
i{\cal L}_{eff}=\left.\frac{d}{ds}\right|_{0}
\frac{\mu^{2s}}{\Gamma(s)}
\int\limits_{0}^{\infty} d (it) (it)^{s-1}
 e^{-iM^2t}\frac{-i}{(4\pi t)^2}
\frac{Bt}{\sin(Bt)}[4\cos^2(Bt)-4\cos(Bt)].
\label{14}
\end{eqnarray}
Furthermore, it is possible to show using (\ref{10})
(or (\ref{A1}) in the case of the electromagnetic field of 
a general form) that
\begin{eqnarray}
\int d^4 x d^4 x^{\prime} 
\langle x |
         e^{iD_{+}^{2}t_1} 
                          | x^{\prime} \rangle
\langle x^{\prime} |
         \not{\!\!D}_{+}e^{iD_{+}^{2}t_2} 
                          | x \rangle = 0. \label{15} 
\end{eqnarray} 

We first note that the term in the brackets in (\ref{14}) contains no term
below order $t^2$ when expanded in powers of $t$; consequently there is no
mass renormalization in the theory as expected due to the supersymmetry of
the model (see also \cite{9703147} and references therein). 

To continue, it is convenient to rewrite ${\cal L}_{eff}$ in  
(\ref{14}) as
\begin{eqnarray}
{\cal L}_{eff}=\left.\frac{d}{ds}\right|_{0}
\frac{\mu^{2s}}{\Gamma(s)} \frac{1}{4\pi^2}
\int\limits_{0}^{\infty} d (it) (it)^{s-3}
 e^{-iM^2t}\frac{Bt}{\sin(Bt)}
\left[-\frac{Bt}{2}\sin(Bt)+\left(\cos^2(Bt)-\cos(Bt)+\frac{Bt}{2}\sin(Bt)
\right)\right].
\label{16}
\end{eqnarray}
Initially, we compute
\begin{eqnarray}
\left.\frac{d}{ds}\right|_{0}
\frac{\mu^{2s}}{\Gamma(s)} \frac{1}{4\pi^2}
\int\limits_{0}^{\infty} d (it) (it)^{s-3}
 e^{-iM^2t}\frac{Bt}{\sin(Bt)}
\left[-\frac{Bt}{2}\sin(Bt)\right]
=\frac{B^2}{8\pi^2}
\left[-\ln\sigma-\ln\frac{|B|}{\mu^2}\right],
\label{17}
\end{eqnarray}
where $\sigma\equiv M^2/|B|=2g^2f^{*}f/|B|$. The remaining integral 
in (\ref{16}) is free of any divergence
at $s=0$, so we are left with
\begin{eqnarray}
{\cal L}_{eff} = - \frac{B^2}{4\pi^2}
\left[\frac{1}{2}\left(\ln\sigma+\ln\frac{|B|}{\mu^2} \right)
+ \int\limits_{0}^{\infty}\frac{dt}{t^2}e^{-it\sigma}
\left([t-\sin t] + [\tan \frac{t}{2} - \frac{t}{2} ]\right)
\right].
\label{18}
\end{eqnarray}
We first note that, using Eq.3.551.9 of \cite{GR},
\begin{eqnarray}
I_1&=&\int\limits_{0}^{\infty}
     \frac{dt}{t^2} e^{-it\sigma} 
\left[\tan\frac{t}{2} - \frac{t}{2} \right]
=\int\limits_{\sigma}^{\infty} d z 
\int\limits_{0}^{\infty}
     \frac{dt}{t} e^{-2tz} 
\left[\tanh t - t \right]\nonumber\\
&=&\int\limits_{\sigma}^{\infty} d z 
\left[\ln\frac{z}{2}
+2 \ln\Gamma\left(\frac{z}{2}\right)
-2 \ln\Gamma\left(\frac{z+1}{2}\right)
-\frac{1}{2z}
\right].
\label{19}
\end{eqnarray}
Next, we get the imaginary part,
\begin{eqnarray}
I_2&=&-i\int\limits_{0}^{\infty}
     \frac{dt}{t^2} \sin(t\sigma) 
\left[t - \sin t\right]
=i\int\limits_{0}^{\infty}dt\left[
\frac{\cos(1-\sigma)t-\cos(1+\sigma)t }{2t^2} 
-\frac{\sin \sigma t}{t}
\right],
\end{eqnarray}
which, upon integrating the first two terms by parts and using 
\begin{eqnarray}
\int\limits_{0}^{\infty}dx
     \frac{\sin \lambda x}{x}= 
\frac{\pi}{2}\frac{\lambda}{|\lambda|}, \quad (\lambda\neq 0)
\end{eqnarray}
becomes
\begin{eqnarray}
I_2 = -i\frac{\pi}{2}(1-\sigma)\theta(1-\sigma).
\label{20}
\end{eqnarray}
Lastly, we have
\begin{eqnarray}
I_3&=&\int\limits_{0}^{\infty}
     \frac{dt}{t^2} \cos(t\sigma) \left[t - \sin t \right]
=\int\limits_{0}^{\infty}dt\left[ \frac{\cos\sigma
t}{t}-\frac{\sin(1-\sigma)t+\sin(1+\sigma)t }{2t^2} \right], 
\end{eqnarray}
which upon integrating the first term by parts becomes 
\begin{eqnarray}
I_3&=&-1 + \lim_{\lambda\to 0}\int\limits_{\lambda}^{\infty}
\frac{dt}{t^2}\left[ \frac{\sin\sigma t}{\sigma}
-\frac{\sin(1-\sigma)t+\sin(1+\sigma)t
}{2} \right] \nonumber\\ &&=-1 + \frac{1+\sigma}{2}\ln(1+\sigma)
+\frac{1-\sigma}{2}\ln|1-\sigma| -\ln\sigma. \label{21} 
\end{eqnarray}
Together, contributions to ${\cal L}_{eff}$ coming from
(\ref{19})--(\ref{21}) show that 
\begin{eqnarray} 
{\cal L}_{eff}&=& -g^2
\frac{B^2}{4\pi^2}\left\{ \frac{1}{2}\ln\frac{2g^2f^*f}{\mu^2} +
U(\sigma)\right\}, \label{22} 
\end{eqnarray} 
with 
\begin{eqnarray}
U(\sigma)&=&\int\limits_{\sigma}^{\infty} d z \left[\ln\frac{z}{2} +2
\ln\Gamma\left(\frac{z}{2}\right) -2 \ln\Gamma\left(\frac{z+1}{2}\right)
-\frac{1}{2z} \right]-\ln(\sigma) \nonumber\\ &&-1 +
\frac{1+\sigma}{2}\ln(1+\sigma) +\frac{1-\sigma}{2}\ln|1-\sigma|
-i\frac{\pi}{2}(1-\sigma)\theta(1-\sigma).\label{us} 
\end{eqnarray}
Including the classical contribution into the effective Lagrangian
(\ref{22}) and trading $\mu$ for the renormalization group invariant scale
\begin{equation} 
\Lambda_\zeta^2=\mu^2e^{-\frac{4\pi^2}{g^2}},\label{rg}
\end{equation} 
we obtain 
\begin{eqnarray} 
V_{B}= \frac{B^2}{8\pi^2}\ln\frac{2g^2 f^*f}{\Lambda_\zeta^2} 
+\frac{B^2}{4\pi^2}U(\sigma). \label{23} 
\end{eqnarray} 
The imaginary part of $U(\sigma)$ arises, as in pure Yang--Mills 
theory, due to unstable (tachyonic) modes in the spectrum of the 
charged vector particle in the presence of a background magnetic field
\cite{BMS,Tseytlin1}. In \cite{NN}, these modes are removed by treating their 
classical part of the action as a Higgs model, i.e. by taking into 
account the quartic self--interaction of these unstable modes
non--perturbatively.  An alternate treatment is given in \cite{F} 
(for a review see \cite{K}). The imaginary part of the effective 
action now disappears, and the real part is believed to remain 
unaltered. Here we must note, that this real part would quite 
likely be shifted though if the coupling between the stable and 
unstable modes were included in the discussion. But even in the 
case of pure Yang--Mills theory, this is a rather complicated
problem which has not been rigorously solved yet. Note however 
that even in the case when this shift turns out to be big, one
should expect that the true vacuum energy is lower then the 
energy of the metastable state with a non-zero imaginary part. 
Other discussions on the stability of translation invariant 
background configurations in Yang--Mills theories are given in 
\cite{L,C}.

In the present model it may appear natural to associate 
the imaginary part with monopole production in an external magnetic 
field. However, perturbation theory does not ``see" these degrees of 
freedom and therefore this interpretation is possibly too far fetched. 

Let us now discuss the vacuum structure predicted by the effective
potential \refs{23}. If we ignore the imaginary part in (\ref{23}), we
find numerically that $d Re V(\sigma)/d\sigma =0$ implies that
$\sigma_{min}\simeq 0.596$ and $U_{min}\simeq -0.358$.  The effective
potential with $\sigma$ at the minimum reads 
\begin{eqnarray}
V_{min}(B) = \frac{1}{2g^2}B^2 
\left(1+\frac{g^2}{4\pi^2}\ln\frac{|B|}{\mu^2}\right)
+\frac{B^2}{4\pi^2}U_{min}+\frac{B^2}{8\pi^2}\ln(\sigma_{min}). 
\label{25}
\end{eqnarray} 
The existence of a negative minimum of $V_{min}(B)$ follows from
the fact that $V_{min}(0)=0$, $V_{min}(B\to\infty) > 0$ and for small $B$,
$V_{min}(B) < 0$ (because of dominance of the logarithmic term in (\ref{25})).
At the minimum, the magnetic field is given by 
\begin{eqnarray} 
B_{min} = \Lambda_\zeta^2\exp \left(-2U_{min}-\frac{1}{2}\right). 
\label{26}
\end{eqnarray} 
However, if $t\es\ha\ln(M^2/\Lambda_\zeta^2)$, then the running
coupling satisfies the equation 
\begin{eqnarray} 
\frac{\hbox{d}}{\hbox{dt}}\bar{g}(t)=-\frac{1}{2\pi^2}g^3(t), 
\label{rg1} 
\end{eqnarray} 
so that
\begin{eqnarray} 
\bar g(t)= \frac{g^2}{1+g^2t/\pi^2}. 
\label{rg2}
\end{eqnarray} 
For $\sigma\es\sigma_{min}$, $B\es B_{min}$ then the value
of $t\es\ha\ln(\sigma_{min}B_{min}/\Lambda^2)$ is such that by \refs{rg1}
the coupling $\bar g^2(t)$ is large and hence the one--loop potential is
unreliable. However, for any value of the external magnetic field, the
scale of the scalar field is fixed, thereby breaking the classical vacuum
degeneracy completely. For large values of $B$, where $\bar g(t)$ is
small, $V_{min}$ in \refs{25} is positive, as expected for a
supersymmetric theory. 

Before seeing how the one-loop approximation can be improved upon, we
turn to $N\es4$ super Yang-Mills theory in the one-loop
approximation.


\section{The N=4 model in a background magnetic field}
\label{four}
The effective potential for SYM$_4$ in a background with constant field 
strength has been considered in \cite{Tseytlin1,Tseytlin2}. As in that 
reference we simplify the algebra by viewing the $N\es 4$ supersymmetric gauge
theory as an N=1 supersymmetric gauge theory in ten dimensions in which
six of the dimensions have been suppressed \cite{Gli}. The original
vector field $A_\alpha^a$ $(\alpha = 1 ... 10)$ decomposes into a four 
component vector field $A_\mu^a$ $(\mu = 1 ... 4)$, three scalars, identified
with $A_5^a ... A_7^a$ and three pseudo-scalars $A_8^a ... A_{10}^a$,
while the original Majorana-Weyl spinor in ten dimensions becomes a set of
four Majorana spinors in four dimensions.  In four dimensions, the
couplings of these matter fields is SU(4) invariant. 

The N=1 supersymmetric gauge theory in ten dimensions that we will
consider has the action
\begin{eqnarray}
S = \int d^{10} x \left[ - \frac{1}{4} G_{\alpha \beta}^a (V) 
G^{a \alpha \beta} (V) 
- \frac{i}{2} \bar{\lambda}^a \not{\!\!D}^{ab} \lambda^b
\right].
\label{30}
\end{eqnarray}

We now use the Honerkamp gauge \cite{Hon}

\begin{eqnarray} 
D^{ab} (A) \cdot Q^b = 0. \label{31} 
\end{eqnarray} 
where V has been decomposed into the sum of a background field $A_\alpha^a$ 
and a quantum field $Q_\alpha^a$.  If the background field in four dimensions
corresponds to a constant background U(1) field $F_{\mu \nu}^a \; (\mu, \;
\nu = 1  ... 4)$ and a constant scalar field of magnitude $gv$, both in the
direction $n^a$ in group space, then 
\begin{eqnarray} 
A_\mu^a = -\frac{1}{2} F_{\mu\nu} \; x^\nu n^a ,
\label{32}
\end{eqnarray} 
\begin{eqnarray} 
\sum_{\alpha=5}^{10} \; A_\alpha^a \; A^{\alpha b} = 
g^2 v^2 n^a \delta^{ab}\equiv M^2 n^a \delta^{ab}. 
\label{33} 
\end{eqnarray} 
The effective action to one-loop order is then given by 
\begin{eqnarray} 
\exp i W^{(1)} = det\left( D^{2 ab}\right)\; det^{-
\frac{1}{2}} (D^{2 ab}g_{\alpha \beta} + 2 f^{apb} F_{\alpha\beta}^p) 
\;det^\frac{1}{2} \left( \not{\!\!D}^{ab} \left( \frac{1+\gamma_{11}}{2}
\right) \right)^2 
\label{34} 
\end{eqnarray} 
in ten dimensions (with the three terms in \refs{34} corresponding to the
contribution of the ghost, vector, and Majorana-Weyl spinor respectively);
dimensionally reducing this to four dimensions with the background field
satisfying the conditions of \refs{32} and \refs{33} converts \refs{34}
into
\begin{eqnarray}
\exp i W^{(1)} &=& det (D^{2 ab} - M^2)\; det^{-\frac{1}{2}} \left(
\left( D^{2 ab} - M^2 \right) g_{\mu\nu} - 2iF_{\mu\nu} \right)
\nonumber \\
& &\left[ det^{-\frac{1}{2}} (D^{2 ab} - M^2) \right]^6 \;
det^{\frac{1}{2}} (D^{2 ab} - M^2 + \frac{1}{2} \sigma_{\mu\nu}
F^{\mu\nu}).
\label{35}
\end{eqnarray}
All derivatives and functional determinants in \refs{35} are understood
to be in four dimensions, all group indices have been suppressed once we
set $(X)^{ab} = i \; f^{apb}(X^p)$, $D = \partial-iA$, and 
$\sigma_{\mu\nu}=\frac{i}{2} [\gamma_\mu , \gamma_\nu]$.

We can now proceed using the techniques outlined in the previous
section.  Regulating as in eq. (11) we see that
\begin{eqnarray}
\left. i W^{(1)} = -\frac{d}{ds}\right|_0 tr \frac{1}{\Gamma(s)} 
\int_0^\infty
\; d i t (it)^{s-1} \; e^{-i [ -D^2 + M^2]t}\;\left[ - \frac{1}{2} \; e^{2F_{\mu\nu} t} \; \; -2 + \frac{1}{2} 
\;
e^{\frac{i}{2} \sigma_{\mu\nu} F^{\mu\nu} t} \right].
\label{36}
\end{eqnarray}
Using (13) and (14), we see that in the presence of an external
magnetic field,
\begin{eqnarray}
\left. i W^{(1)} = \frac{d}{ds}\right|_0 \frac{1}{\Gamma(s)} 
\int_0^\infty \; d \;
it (it)^{s-1} \left( -\frac{i}{(4 \pi t)^2} \frac{Bt}{\sin B t}
\right) (4 \cos^2 B t + 4 - 8 \cos Bt) e^{-iM^2t}.
\label{37}
\end{eqnarray}
(An overall factor of 2 comes from the trace in group space.)  The
integral over $t$ in \refs{37} is free of divergences at $s=0$, and 
hence, upon using some trigonometric identities, we see that 
\begin{eqnarray}
i W^{(1)} = \frac{8 B^2}{(4 \pi)^2} \int_0^\infty d (it) (it)^{-2}
 \left[ \left( \tan \frac{t}{2} - \frac{t}{2} \right) + \left(
\frac{t}{2} - \frac{1}{2} \sin t \right) \right] e^{-i \sigma t},
\label{38}
\end{eqnarray}
where $\sigma = M^2 / B$.  The integrals in \refs{38} are given in 
(22),
(25) and (27) so that
\begin{eqnarray}
i W^{(1)} &=& -\frac{i B^2}{2 \pi^2} \left[ \int_\sigma^\infty  dz 
\left(
\ln \frac{z}{2} + 2 \ln \Gamma \left( \frac{z}{2} \right) - 2 \ln
\Gamma \left( \frac{z+1}{2} \right) - \frac{1}{2z} \right) 
\right.\nonumber
\\
&+& \left. \frac{1}{2} \left( -1 + \frac{1 + \sigma}{2} \ln 
(1+\sigma) +
\frac{1 - \sigma}{2} \ln \left| 1 - \sigma \right| - \ln \sigma 
- i \frac{\pi}{2} (1 - \sigma) \theta (1 - \sigma)
\right) \right] \nonumber \\
& & \equiv -\frac{i B^2}{2 \pi^2} \overline{U} (\sigma). 
\label{39}
\end{eqnarray}
The effective potential to one-loop order is hence given by
\begin{eqnarray}
V_B = \frac{1}{2} B^2 \left[ 1 + \frac{1}{\pi^2} \overline{U}
(\sigma) \right].
\label{40}
\end{eqnarray}
We note in passing that the result \refs{39} could also be obtained 
by identifying the mass term in \cite{Tseytlin2} with the background scalar. 
As for the imaginary part the discussion 
in the last section could be repeated here. In particular for $M\to 0$ 
we recover the result of \cite{Tseytlin2} for a vanishing scalar background. 

The minimum value of ReV$_B$ occurs when $d \left[ Re \overline{U}
(\sigma) \right] / d \sigma = 0$; this occurs when
\begin{eqnarray}
\sigma_{min} = .797
\label{41}
\end{eqnarray}
at which point
\begin{eqnarray}
Re \overline{U}_{min} = -.247
\label{42}
\end{eqnarray}
so that $V_B > 0$ at $\sigma = \sigma_{min}$, in agreement with the
theorem that the vacuum energy of a supersymmetric theory is non-negative. 
However, for $B \neq 0$, the degeneracy in the vacuum expectation value of
the scalar particle is broken by \refs{41}.  This result is reliable in
perturbation theory, for in $N = 4$ SUSY the coupling $g$ does not run and
hence may be chosen to be small irrespective of the value of B. Furthermore 
the non-renormalization theorem of \cite{Dine} excludes further corrections to 
the $B^4$ term.

\section{Non-perturbative corrections}
\label{five}

In this section we analyze the effect of non-perturbative corrections to 
the effective potential in two different ways. First by including 
all instanton corrections to the running coupling and second by 
evaluation the effective potential within the dual description. 
 
\subsection{Instanton improved potential for N=2}

We start by
comparing \refs{23} with the exact low energy effective action \cite{SW}. 
\begin{eqnarray}
{\cal L}_{SW} = \frac{1}{4\pi}Im \left[\int d^4\theta 
\frac{\partial {\cal F}(A)}{\partial A} \bar{A} + 
\int d^2\theta 
\frac{1}{2}\frac{\partial^2 {\cal F}(A)}{\partial A^2} 
W_{\alpha}W^{\alpha}\right],
\label{SW1}
\end{eqnarray}
where $A\es a+\theta\chi+\cdots$ and $W_{\alpha}\es \chi_\al+\cdots$ are
the $N=1$ chiral and vector multiplets, respectively. (The scalar
component of $A$ is proportional to $f$ in our notation.) The function
${\cal F}(A)$ in the region of validity of perturbation theory ($|a|\to
\infty$) is given by \cite{SW,Sei}
\begin{eqnarray}
{\cal F}(A)= \frac{i}{2\pi}A^2\left(\ln 
\frac{A^2}{\Lambda^2}+c\right),
\label{SW2}
\end{eqnarray}
where $c$ depends on the renormalization scheme. Matching the scales as in
\cite{Pouliot} we get from \refs{SW2} with $\Lambda^2_\zeta\es
\Lambda^2_{\overline {DR}}\es\ha \Lambda^2$ (here $\Lambda$ corresponds to
the scheme used in \cite{SW} and $\Lambda_{\overline {DR}}$ is the scale
of dimensional reduction combined with minimal subtraction)
\begin{eqnarray}
V_{B}\simeq\frac{B^2}{8\pi^2}\ln\frac{2|a|^2}{\Lambda_\zeta^2},
\quad \mbox{as}\quad  \frac{|a|}{\Lambda_\zeta}\to \infty,
\label{24}
\end{eqnarray}
with $c=0$. From \refs{us} it follows that $U(\sigma)\rightarrow 0$ for
$\sigma\rightarrow\infty$, so that \refs{24} is identical with (\ref{23})
if we make the identification $a\es gf$. This is also the identification
which is consistent with BPS mass formula $M^2\es 2|an_e+a_Dn_m|^2$. 

The analysis of \cite{SW} was based on the following argument:  each fixed
value of the scalar field $|a|$ defines a distinct vacuum of the system;
different values of $|a|$ define inequivalent vacua. The reason for this
(at least in the asymptotic region $|a|\to\infty$) is well understood. As
is seen from (\ref{24}), for any value of $|a|$ the minimum of the
effective potential is achieved by taking $B=0$. Since at this minimum,
the value of the potential is zero, the lowest possible in supersymmetric
models, we conclude that there exist different vacua for each choice of
the value of $|a|$. On the other hand, for values of $|a|$ of the order of
$\Lambda$ the potential (\ref{24}) is unbounded below. It was argued in
\cite{SW} and later shown \cite{USW} that the function ${\cal F}(A)$ has a
unique extension to the strong-coupling limit, compatible with
supersymmetry and a finite number of singularities. Furthermore the
quantum moduli-space is in one-to-one correspondence with the parameter
$u=\tr\<\phi^2\>$ which takes its value in the upper half plane
\cite{USW}. For large values of $a$, $u = \frac{1}{2} a^2$ since in this
region $\phi = \frac{1}{2} a \sigma^3$.  For $u\simeq \Lambda^2$ the
effective coupling diverges due to the appearance of massless composite
fields which are magnetic monopoles. In the neighbourhood of this
singularity the theory should then be accurately described by a dual
theory which is magnetic $N=2\;  QED$\footnote{ For the $N\es 4$ model
discussed in the last section the dual theory would be itself.} \cite{SW}. 

From the above discussion we draw the following conclusions. First of
all, the one-loop result \refs{23} can be improved by replacing the first
term by the corresponding non-perturbative expression \cite{SW}. The
higher loop and non-perturbative contributions to $U(\sigma)$ which are
important in the regime where the scalar and/or the magnetic field are of
the order of $\Lambda$, will be approximated by computing the analogue of
$U(\sigma)$ in the corresponding dual model. This will be done by
evaluating to one--loop order the effective potential in $N\es2\;  QED$ in
the presence of background scalar and electromagnetic fields which are
dual to the fields appearing in \refs{23}. The contribution to the
analogue of the first term in \refs{23} can be compared to the
non-perturbative expression of this first term in \refs{23}, obtained by
using the methods of \cite{SW}.  This complements the comparison of the
non-perturbative extension to the instanton contribution to the effective
potential of $N\es2$ $SU(2)$ super-Yang-Mills theory (see [27]). 
Furthermore the remaining part of the one--loop effective potential in
$N\es 2\; QED$ should approximate the non-perturbative extension of
the function $U(\sigma)$ in \refs{23}.  The accuracy of this approximation
relies on to what extent duality in $N\es 2$ Yang-Mills theory is realized
away from the strict vacuum and the strong--weak coupling singularities.
We therefore expect it to be good for small magnetic fields in the
neighbourhood of the point where monopoles are massless, while nothing is
known in the general case.  We cannot test this directly as higher loop
and instanton corrections to the effective potential in the presence of a
background magnetic field are unknown. 

Let us first implement the instanton corrections. For this we substitute
the exact result \cite{SW} for the running coupling in the first term in
\refs{23}, which then becomes
\begin{eqnarray}
V_{B}= \frac{B^2}{8\pi}\I\left(\tau(u)\right),\label{23i}
\end{eqnarray}
where the explicit expression for $\tau$ and $f$ as a 
function of $u\es \<tr\phi^2\>$ 
are given by \cite{SW,Bilal} 
\begin{equation}
\tau(u)=i\frac{F[1/2,1/2,1,\frac{u-1}{u+1}]}{F[1/2,1/2,1,\frac{2}{u+1}]}.
\end{equation}
This is the full non-perturbative expression for the effective potential
in the presence of a very weak magnetic field. Concerning the second term
in \refs{23} there appears to be an ambiguity. There are two different
non-perturbative expressions for $\sigma$, each of which reduce to the
perturbative expression (below \refs{38}); both 
$\sigma\es \frac{2|a|^2}{|B|}$ and the gauge-invariant form $\sigma\es
\frac{4u}{|B|}$ reduce to the perturbative expression.  In Fig.~1  
the improved effective potential is plotted for different values 
of $u$ (horizontal) and the magnetic field $|B|$ for both possible
parametrizations of $\sigma$. This shows that the qualitative behaviour 
is the same for both choices. The exact relation between $u$ and 
$a$ is \cite{SW}
\begin{equation}
a(u) = \sqrt{2} \sqrt{u+1}\; F \left( -\frac{1}{2}, \frac{1}{2}, 1,
\frac{2}{1+u} \right).
\end{equation}

\subsection{ Dual description}

The only theory with the same number of degrees of freedom as the
YM-theory, which has $N\es2$ SUSY and in which the coupling runs to zero
at small scale is $N\es2$ SUSY QED with magnetic rather than electric
charges. In component form its Lagrangian reads \cite{West}
\begin{eqnarray}
S&=&\int d^4 x \left\{-\frac{1}{4}
F_{\mu\nu}F^{\mu\nu}
-(\pa_\mu\phi_D)^{*}(\pa^\mu\phi_D)
-\bar{\lambda}^i\pa\!\!\!/\lambda_{i} +\ha X^2\right. 
\nonumber \\
&&\qquad\qquad-(D_\mu A^i)^*D^\mu A_i-\bar{\psi}\not{\!\!D}\psi 
+|F^i|^2\nonumber \\
&&\qquad\qquad-i2g\bar\lambda^i\psi\bar A_i+i\sqrt{2}g\bar\psi
  [\phi_D\gamma_{-}+\phi_D^{*}\gamma_{+}]\psi\nonumber \\
&&\left.\qquad\qquad+4igX^{ij}\bar A_i A_j-2g^2 |\phi_D|^2
\bar A^i A_i\right\}.
\label{d1}
\end{eqnarray}
The background configurations are now $\phi_D \es f_D$ and 
$F_{D\mu\nu}$ respectively. We follow \cite{SW,FS} in order to 
determine $F_{D\mu\nu}$. Consider 
\begin{equation}\label{F_D}
S=\frac{1}{16\pi}\int\d^4 x\left[-\frac{4\pi}{\bar g^2}F_{\mu\nu}
F^{\mu\nu}-\frac{\bar\theta}{2\pi}F_{\mu\nu}\tilde 
F^{\mu\nu} -4V^\mu\pa^\nu\tilde F_{\mu\nu}\right],
\end{equation}
where $\bar g,\bar\theta$ denote the effective coupling and vacuum 
angle, $\tilde F_{\mu\nu} \es \ha \epsilon_{\mu\nu\lambda\sigma}
F^{\lambda\sigma}$   and $V^\mu$ is a Lagrange multiplier vector 
field imposing the  Bianchi  identity $\pa_\nu\tilde F^{\mu\nu}\es 0$. 
Varying \refs{F_D} with  respect to  $F_{\mu\nu}$ then leads to 
\begin{equation}\label{FD2}
F_{D\mu\nu}=\frac{4\pi}{\bar g^2}\tilde 
F_{\mu\nu}-\frac{\bar\theta}{2\pi}
F_{\mu\nu},
\end{equation}
where $F_{D\mu\nu}=\pa_\mu V_\nu-\pa_\nu V_\mu$ is the field strength 
in the dual theory. This is consistent with 
\begin{equation}\label{FMU}
F_{\mu\nu}=\frac{4\pi}{\bar g_D^2}\tilde 
F_{D\mu\nu}-\frac{\bar\theta_D}{2\pi}F_{D\mu\nu},
\end{equation}
provided $\tau_D\es \frac{\bar\theta_D}{2\pi}+\frac{i4\pi}{\bar 
g_D^2} 
\es -\frac{1}{\tau}$. 
With $\epsilon_{0123}\es 1$ and $F_{12}\es B$ we see by \refs{FD2} 
that 
\begin{equation}\label{FD03}
F_{D03}=\frac{4\pi}{\bar g^2}B\mtx{and} 
F_{D12}=-\frac{\bar\theta}{2\pi}B.
\end{equation}
The one-loop effective potential for the action \refs{d1} is then 
given by 
the following analogue of (16)
\begin{equation}\label{wd1}
iW^{(1)}_D=-2\tr\ln(D^2_{D_+}-M_D^2)+\tr 
\ln(i\di_D+\sqrt{2}gf_D\gamma_+\sqrt{2}g\bar
f_D\gamma_-),
\end{equation}
where
\begin{equation}
D_{D+}^2=(\pa+iV)^2\mtx{and}M_D^2=2g_D^2\bar f_Df_D,
\end{equation}
with $g_D$ being the ``classical coupling" of the dual $QED$. Note the 
presence of the chiral
mass term in the Dirac operator. It leads to a phase dependence of 
the Dirac determinant of the
form of the chiral anomaly proportional to
$\theta_D \int 
F_{D\mu\nu}\tilde F^{\mu\nu}_D$. However, this term, being linear in 
the 
electric field, drops out in the effective potential. We can 
therefore 
ignore this phase and replace the last term in \refs{wd1} by 
\begin{equation}
\ha
\tr\ln(D_{D+}^2+\ha\sigma^{\mu\nu}F_{D\mu\nu}-M_D^2).
\end{equation}
To continue we use the identity \refs{A8}  
\begin{equation}
\tr\; 
e^{\frac{i}{2}F_{D}^{\mu\nu}\sigma_{\mu\nu}t}=4\cosh(tK_+)\cos(tK_-),
\end{equation}
where by \refs{FD03}, \refs{A2}
\begin{equation}
K_+=E_D=\frac{4\pi}{\bar g^2}B = \hbox{Im}[\tau]B\mtx{and} 
K_-=-B_D=\frac{\bar\theta}{2\pi}B = \hbox{Re}[\tau]B.
\end{equation}
The steps which lead from \refs{-16} to \refs{16} can now be repeated 
to give 
\begin{eqnarray}\label{gd1}
iW^{(1)}_D=\left.\frac{d}{ds}\right|_{0}
\frac{\mu^{2s}}{\Gamma(s)}\int\limits_0^\infty \d
it(it)^{s-1} e^{-i
M_D^2t}\left(\frac{-i}{(4\pi t)^2}\right)\frac{E_Dt}{\sinh 
E_Dt}\frac{B_Dt}{\sin
B_Dt}\left[2-2\cosh E_Dt\;\cos B_Dt\right] .
\end{eqnarray}
Let us first have a closer look at the leading term in the magnetic 
field $B$. In leading order \refs{gd1} simplifies to 
\begin{eqnarray}\label{gd2}
iW^{(1)}_D=\left.\frac{d}{ds}\right|_{0}\frac{\mu^{2s}}{\Gamma(s)}\int
\limits_0^\infty \d (it)(it)^{s-1} e^{-i M_D^2t}
\left(\frac{-i}{(4\pi t)^2}\right)\cdot\left[\left(\frac{\bar\theta}{2\pi}
Bt\right)^2-\left(\frac{4\pi}{\bar g^2}Bt
\right)^2\right].
\end{eqnarray}
Performing the remaining integration and taking the Legendre transform 
with respect to $E_D$ this leads to the dual effective 
potential
\begin{eqnarray}\label{vv1}
V_D(B,f_D) &=& 
-\frac{1}{(4\pi)^2}\ln\left(\frac{M_D^2}{\mu^2}\right)
\left(\left(\frac{4\pi}{\bar g^2}\right)^2
- \left(\frac{\bar\theta}{2\pi}\right)^2\right)B^2 \nonumber\\
&=&  - \frac{1}{(4\pi^2)} \ln \left( \frac{M_D^2}{\mu^2} \right)
\left[ E_D^2 - B_D^2 \right]
\end{eqnarray}
Now, using the BPS mass formula for a minimally charged monopole 
$M^2\es 2|a_D|^2$ we identify $a_D\es g_Df_D$. Then, using the exact 
expression \cite{SW} for $\tau_D(a_D) = -\tau^{-1} (a)$ we can 
rewrite \refs{vv1} as 
\begin{equation}\label{vv2}
V_D(B,f_D)=\frac{1}{8\pi}\hbox{Im}[\tau_D]|\tau|^2B^2=\frac{1}{8\pi}
\hbox{Im}[\tau]B^2,
\end{equation}
since $\tau_D (a_D) \approx \frac{i}{\pi} ln \left( \frac{a_D}{\pi}
\right)$ in the region where (67) is valid, showing that to leading 
order in $B$ the dual potential is identical  with  the original 
potential as it must be in order to be consistent with \cite{SW}. 
In this limit the duality invariance of the effective potential is 
easy to establish \cite{FS}.

The full expression for the dual one--loop effective action 
\refs{gd1} does not appear to be easily tractable. A simplification 
occurs, however, if we take $\bar\theta\es 0$. This is consistent as 
long as the moduli parameter $u$ takes values on the real axis with
$u>\Lambda^2$.  In that  situation \refs{gd1} takes the form 
\begin{eqnarray}\label{gd12}
iW^{(1)}_D&=&\left.\frac{d}{ds}\right|_{0}
\frac{\mu^{2s}}{\Gamma(s)}\int\limits_0^\infty \d (it)(it)^{s-1}
e^{-i M_D^2t}\left(\frac{-i}{(4\pi t)^2}\right)\frac{E_Dt}
{\sinh E_Dt}\left[2-2\cosh E_Dt\right]\nonumber\\
&=&-i\frac{E_D^2}{8\pi^2}\ln\left(\frac{M_D^2}{\mu^2}
\right)+i\frac{E_D^2}{8\pi^2}\int\limits_0^\infty \d tt^{-2}e^{-i 
\sigma_Dt}
\left[\tanh\left(\frac{t}{2}\right)-\frac{t}{2}\right],
\end{eqnarray}
where $\sigma_D\es M_D^2/E_D$. The corresponding effective potential 
is obtained, as usual, via the Legendre transform. Taking the real 
part we have  
\begin{eqnarray}\label{gd13}
V_D[u,E_D]&=&\frac{\pa W_D}{\pa E_D}E_D-W_D\nonumber\\
&=&\frac{1}{8\pi}
\hbox{Im}[\tau]B^2 
+\frac{E_D^2}{8\pi^2}\left(1-\sigma_D\frac{\pa}{\pa\sigma_D}\right)
\int\limits_0^\infty \d tt^{-2}e^{-i \sigma_Dt}
\left[\tanh\left(\frac{t}{2}\right)-\frac{t}{2}\right]\nn\\
&=&W_D+\frac{E_D^2}{8\pi^2}\sigma_D\int\limits_0^\infty \frac{\d t}
{t}\sin(\sigma_Dt)\left[\tanh\left(\frac{t}{2}\right)-\frac{t}{2}\right],
\end{eqnarray}
where we have used  $E_D\frac{\pa}{\pa E_D}\es -\sigma_D\frac{\pa}
{\pa \sigma_D}$ and we  have again substituted the exact expression 
for the leading term in  $E_D$. To continue we use  
\begin{eqnarray}\label{gd14}
&\hbox{Re}&\int\limits_0^\infty \d tt^{-2}e^{-i \sigma_Dt}
\left[\tanh\left(\frac{t}{2}\right)-\frac{t}{2}\right]
\nonumber\\&=&\ha\cos(2\sigma_D)
\left[si(2\sigma_D)-ci(2\sigma_D)\right]+\int\limits_0^\infty \d t
t^{-2}\cos(\sigma_D t) \left[\tanh\left(\frac{t}{2}\right)
-\frac{t}{2}\right],
\end{eqnarray}
where $si(z)$ and $ci(z)$ are the integral sine and cosine 
respectively. The remaining part can be calculated numerically. 
The resulting effective potential is plotted in Fig.~2.

It is interesting to isolate the contribution to the effective 
potential which comes from the non-leading terms only. The leading 
terms ($O(B^2)$) are, of course, identical because they are exact and the 
exact effective potential is duality invariant. The terms 
of order $B^4$ can obtained by expanding \refs{18} and \refs{gd12}, leading to 
\begin{eqnarray}
V(B)|_{B^4}&=&-\frac{5}{2^{6}\pi^2 3!}\frac{B^4}{|a|^4}\mtx{and}\nonumber\\
V_D(B)|_{B^4}&=&\frac{5}{2^{8}\pi^2}\frac{\hbox{Im}[\tau] B^4}{|a_D|^4},
\end{eqnarray}
respectively. The difference in sign is consistent with the absence 
of a non-trivial minimum in the dual description. 
The complete non-leading contributions to the effective- 
and dual effective  potentials are  plotted in Fig.~3. 
Note that up to a global sign they are almost identical.

\section*{Discussion}
\label{six}
In this paper we have analyzed the effective potential for 
$N\es 2$ and $N\es 4$ SUSY Yang-Mills theory within different 
approximations. Our main finding is that the non-trivial 
minimum that appears generically in one-loop approximations 
survives even if the leading order in the background 
magnetic field is evaluated exactely, but is absent in the dual 
description which takes into account the monopole dynamics. This 
gives support to the idea that monopoles stabilize the 
theory in the strongly coupled regime. It would be of 
interest to know whether this qualitative feature survives in 
a non-supersymmetric theory. 

An implicit assumption in our analysis is that the simplest form of 
duality proposed in \cite{SW} is approximately realized at least for 
small but non-vanishing magnetic fields. Our results appear to 
be consistent with this assumption. Furthermore, the combination 
of perturbative Yang-Mills- and dual effective potential leads to 
a self consistent effective potential (i.e. compatible with the 
symmetries of the theory) for all values of the external field. 

The leading order contribution in the background magnetic field 
to the effective potential being evaluated 
exactely, the difference between the effective potential 
in the fundamental- and dual description is due to non-leading 
contributions. We find that up to an overall sign these contributions 
are almost identical in the two description. At present it 
is not clear to us whether this could be anticipated.


\section*{Acknowledgements}

We would like to thank C. Ford for collaboration in the early stage of this 
work and A. Tseytlin for drawing our attention to refs. 
\cite{Tseytlin1,Tseytlin2}. NSERC is acknowledged for financial support. 
R. and D. MacKenzie were helpful in motivating this 
research. I.S. would like to thank the Department of Applied 
Mathematics at University of Western Ontario in London 
for hospitality during the  first stage of this project. 
The work of I.A.S. was supported by the U.S. Department 
of Energy Grant \#DE-FG02-84ER40153. Michael Haslam helped 
with the computer evaluations of
$\sigma_{min}$.

\appendix


\section{General Case of the Electromagnetic Field}

Using the result of \cite{Sch} for the matrix elements 
of interest, and the method of \cite{BS} dealing with 
functions of matrix argument $F_{\mu\nu}$, we obtain
\begin{eqnarray}
\langle x^{\prime} | \exp ( iD_{+}^{2}t ) | x \rangle
&=&\frac{-i}{(4\pi t)^2} 
\frac{tK_{-}}{\sin(tK_{-})}\frac{tK_{+}}{\sinh(tK_{+})}
\nonumber\\
&\times& \exp \left[ 
\frac{i}{4}(x-x^{\prime})_\mu C^{\mu\nu} (x-x^{\prime})_\nu 
 + i \int\limits_{x^{\prime}}^{x} A_{\lambda} d z^{\lambda}
 \right]
\label{A1}
\end{eqnarray}
where the integral over $z$ is taken along the straight 
line running from $x^{\prime}$ to $x$. In (\ref{A1}), 
we also introduced two independent invariants:
\begin{eqnarray}
K_{\pm}=\sqrt{ \sqrt{ {\cal F}^2+{\cal G}^2}\pm{\cal F}}, 
\quad \mbox{with}
\quad  {\cal F}=-\frac{1}{4}F_{\mu\nu}F^{\mu\nu}, 
\quad  {\cal G}=\frac{1}{8}\epsilon^{\alpha\beta\mu\nu}
                F_{\alpha\beta}F_{\mu\nu}
\end{eqnarray}
and
the following matrix:
\begin{eqnarray}
C^{\mu\nu}&=& \frac{g^{\mu\nu}K_{-}K_{+}}{K_{+}^2+ K_{-}^2}
\left[\frac{K_{+}}{\tan tK_{-} } 
+ \frac{K_{-}}{\tanh tK_{+} }  \right] 
-\frac{(F^2)^{\mu\nu}}{K_{+}^2+ K_{-}^2}
\left[\frac{K_{+}}{\tanh tK_{+} } 
- \frac{K_{-}}{\tan tK_{-} }  \right]
\end{eqnarray}

When dealing with propagators for fermions, one also needs
a convenient expression for the following matrix
\begin{eqnarray}
f(t)&=& \exp \left(-\frac{i}{2}F_{\mu\nu}\sigma^{\mu\nu} t \right),
\quad \sigma^{\mu\nu}=\frac{i}{2} [\gamma^\mu,\gamma^\nu]
\label{A4}
\end{eqnarray}
which appears in \cite{Sch} only in this awkward form. 
A representation with explicit Dirac matrix 
structure was presented in \cite{GCZ}. Below, we derive another
representation which has an explicit structure in 
both Dirac and tensor indices. 

It is easy to see that
\begin{eqnarray}
f^{\prime\prime}(t)&=& 
-\frac{1}{4}F_{\mu\nu}\sigma^{\mu\nu} 
F_{\alpha\beta}\sigma^{\alpha\beta}f(t)= 
2({\cal F} + i\gamma^5{\cal G}) f(t)=
     (K_{+} + i\gamma^5K_{-})^2 f(t)
\label{A2}
\end{eqnarray}
(Here, we have used the following notation:
$\gamma^5=-i\gamma^0\gamma^1\gamma^2\gamma^3  $ and
$\epsilon^{0123}=+1$.)

Solving the homogeneous differential equation (\ref{A2}) gives
\begin{eqnarray}
f(t)&=& C_1 \cosh[(K_{+} + i\gamma^5K_{-})t] 
       +C_2 \sinh[(K_{+} + i\gamma^5K_{-})t]
\end{eqnarray}
with $C_1=1$ and 
\begin{eqnarray}
C_2= - \frac{i}{2(K_{+}^2+K_{-}^2)}(K_{+} - i\gamma^5K_{-}) 
F_{\mu\nu}\sigma^{\mu\nu} ,
\end{eqnarray}
in order to satisfy the conditions $f(0)=1$, 
$f^{\prime}(0)=-iF_{\mu\nu}\sigma^{\mu\nu}/2$.
Thus, we obtain a closed form expression for (\ref{A4}):
\begin{eqnarray}
&& \exp \left(-\frac{i}{2}F_{\mu\nu}\sigma^{\mu\nu} t \right)=
\cosh(tK_+)\cos(tK_-)
       \Bigg[ 1 - i\gamma^5 \tanh(tK_+)\tan(tK_-)+
  \nonumber\\
&&- \frac{iF_{\mu\nu}\sigma^{\mu\nu}}{2(K_{+}^2+K_{-}^2)}
    \left(K_+\tanh(tK_+)+K_-\tan(tK_-)\right)+
  \nonumber\\
&& - \gamma^5\frac{F_{\mu\nu}\sigma^{\mu\nu}}
{2(K_{+}^2+K_{-}^2)}
    \left(K_-\tanh(tK_+)-K_+\tan(tK_-)\right)
\Bigg]
\label{A8}
\end{eqnarray}
This relation can also be used to analyze the properties 
of a Dirac spinor under a  Lorentz transformation.

In the case of propagators for vector fields, 
one needs a closed expression for the spin factor 
$\exp (-2Ft)_{\mu\nu}$ (in the $\xi=1$ gauge). Again using 
the method of \cite{BS}, we obtain:
\begin{eqnarray}
&&\exp (-2Ft)_{\mu\nu}=\frac{1}{(K_{+}^2+K_{-}^2)}
\Bigg\{
+g_{\mu\nu}\left[K_{-}^2 \cosh(2tK_{+})  
                 + K_{+}^2\cos(2tK_{-})  \right]
\nonumber \\
&&+(F^2)_{\mu\nu}\left[\cosh(2tK_{+}) 
                    - \cos(2tK_{-}) \right]
- F_{\mu\nu}\left[ K_{+} \sinh(2tK_{+})
                 +  K_{-} \sin(2tK_{-})  \right]
\nonumber\\
&&+ F_{\mu\nu}^{*}\left[K_{-} \sinh (2tK_{+})  
                 - K_{+} \sin(2tK_{-})  \right]
\Bigg\}
\label{A9}
\end{eqnarray}  

Using the proper time representation (\ref{8}), (\ref{9})
and equations (\ref{A1}), (\ref{A8}) and (\ref{A9}),
one can obtain convenient expressions for the propagators
of charged scalar, fermion and vector fields appearing in 
(\ref{7}).


\newpage
\iffigs

\begin{figure}[h]
\epsfxsize=6cm
\hbox{
\epsfbox{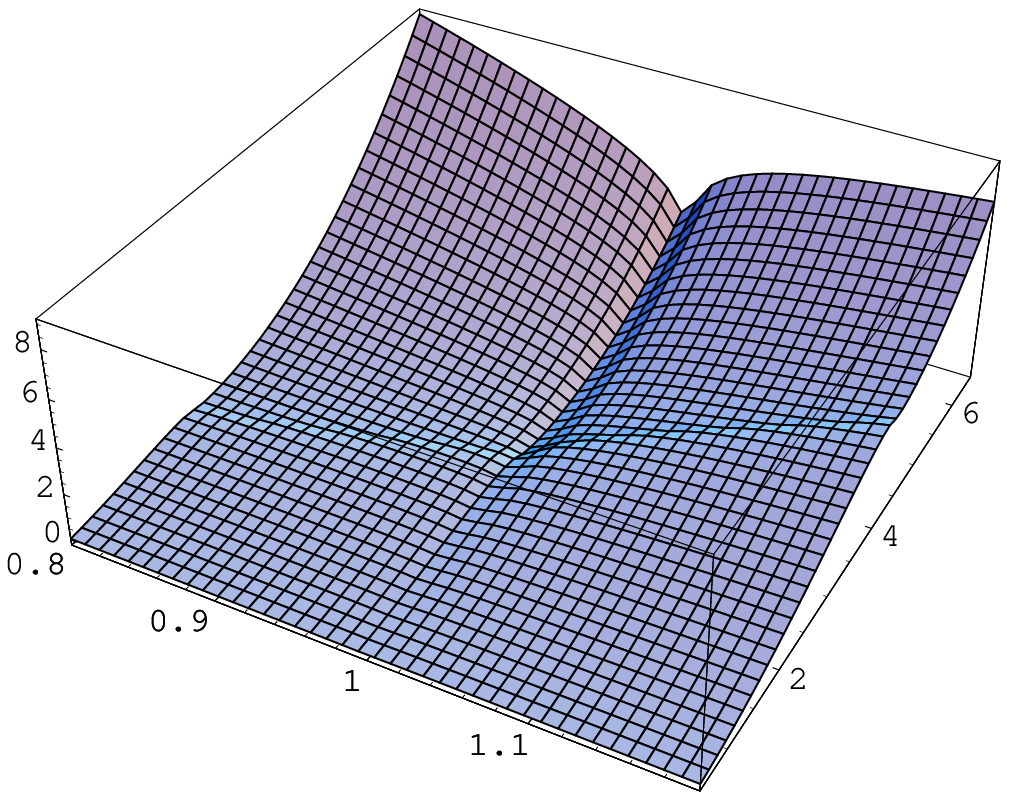}\hspace{3cm}
\epsfxsize=6cm
\epsfbox{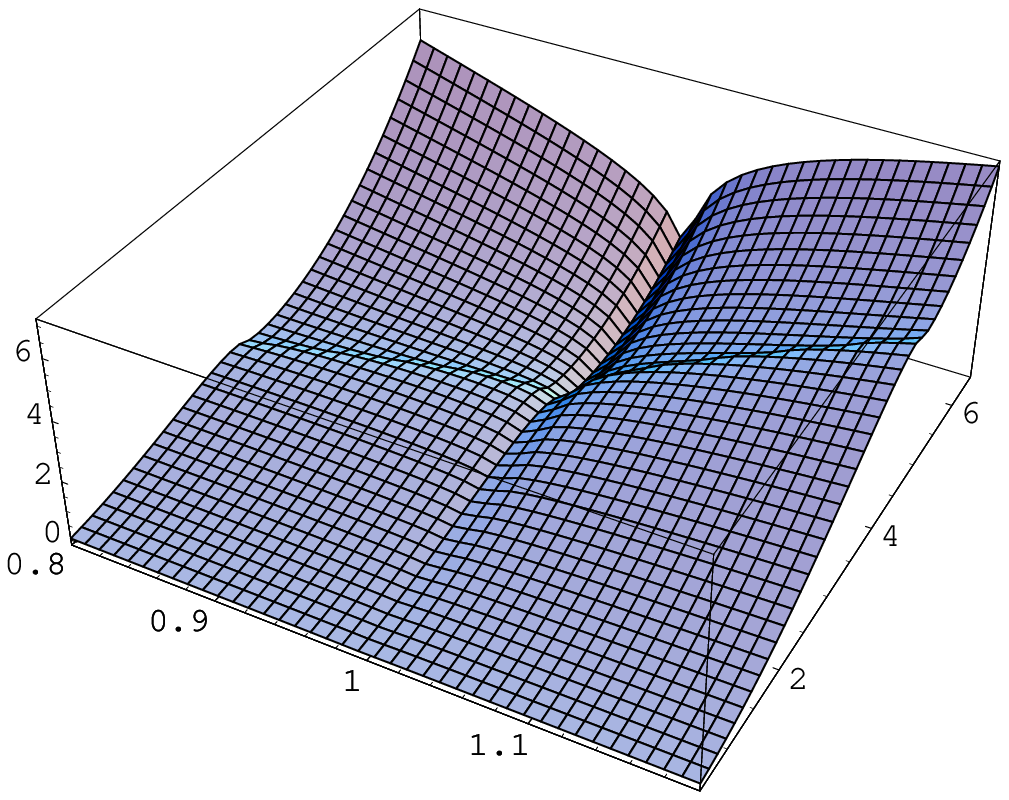}
}
\caption{Effective potential: $V(\frac{a[u]}{\Lambda},B)$ and 
$V(\frac{u}{\Lambda^2},B)$ as a function of $u=\tr\<\phi^2\>$ 
and $B$}
\end{figure}
\vspace{1cm}

\begin{figure}[h]
\epsfxsize=6cm
\epsffile{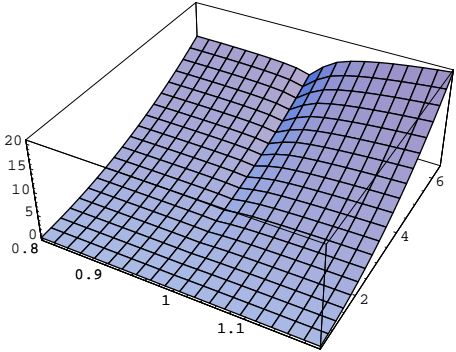}
\caption{Dual Effective potential: 
$V_D(\frac{a_D[u]}{\Lambda},B)$ as a function of $u$ and $B$}
\end{figure}
\vspace{1cm}

\begin{figure}[h]
\epsfxsize=6cm
\hbox{
\epsfbox{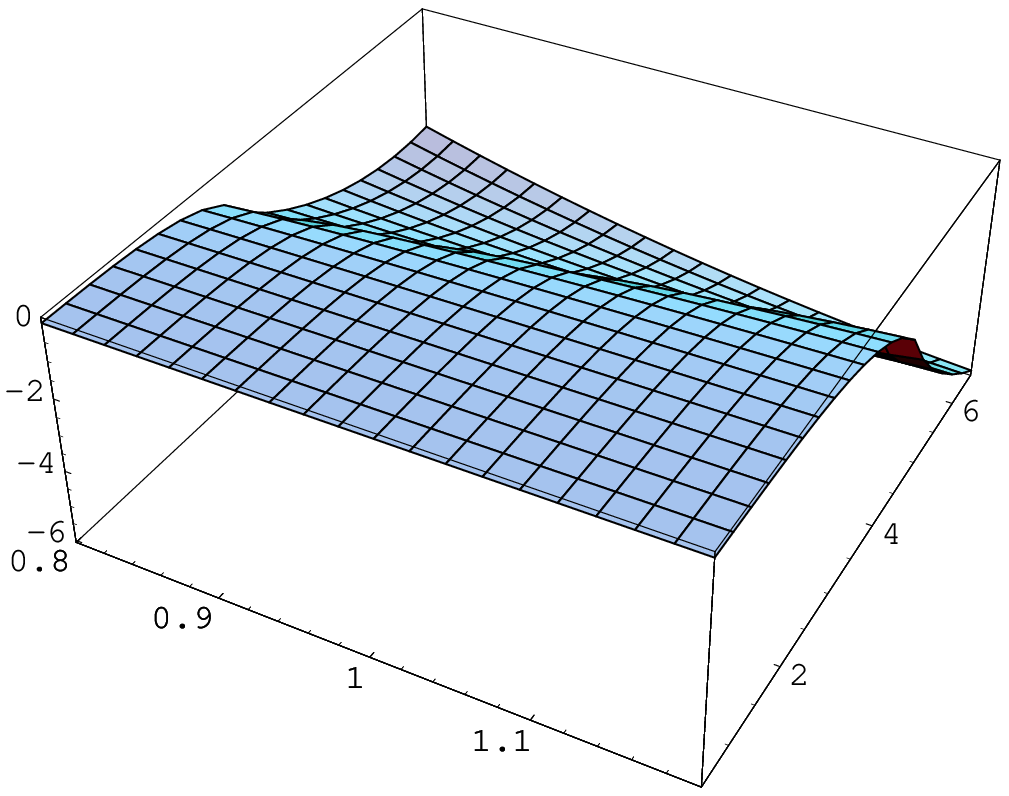}\hspace{3cm}
\epsfxsize=6cm
\epsfbox{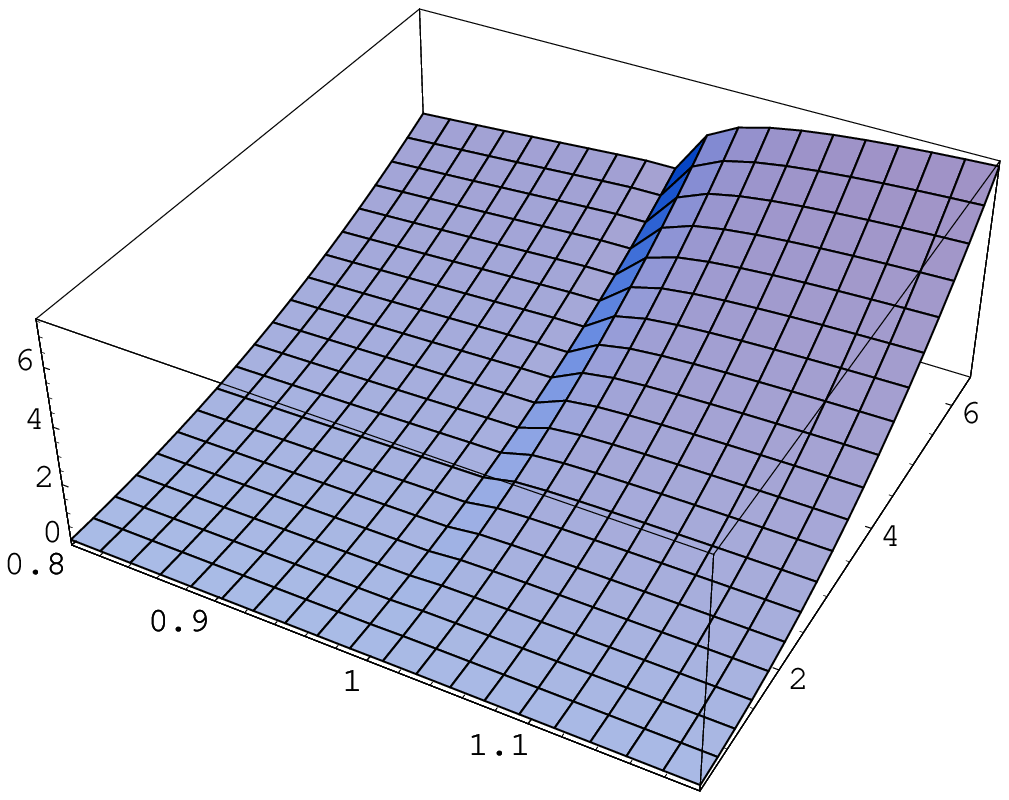}
}
\caption{Contribution to the effective- and dual effective potential 
from the non-leading terms as a function of $u$ and $B$ }
\end{figure}
\vspace{1cm}

\else
\message{No figures will be included. See TeX file for more
information.}
\fi

\end{document}
\end